\documentclass[10pt,journal,compsoc]{IEEEtran}
\ifCLASSOPTIONcompsoc
  \usepackage[nocompress]{cite}
\else
  \usepackage{cite}
\fi
\ifCLASSINFOpdf
\else
\fi
\IEEEoverridecommandlockouts
\usepackage{cite}
\usepackage{amsmath,amssymb,amsfonts}
\usepackage{algorithmic}
\usepackage{graphicx}
\usepackage{textcomp}
\usepackage{xcolor}
\usepackage{diagbox}
\usepackage{multirow}
\usepackage{comment}
\usepackage{caption}
\usepackage{subcaption}

\def\BibTeX{{\rm B\kern-.05em{\sc i\kern-.025em b}\kern-.08em
		T\kern-.1667em\lower.7ex\hbox{E}\kern-.125emX}}

\newsavebox{\ieeealgbox}

\begin{document}

\title{Bridging the Gap between Deep Learning and Frustrated Quantum Spin System for Extreme-scale Simulations on New Generation of Sunway Supercomputer}

\author{\IEEEauthorblockN{Mingfan Li\IEEEauthorrefmark{2}, Junshi Chen\IEEEauthorrefmark{2}, Qian Xiao, Fei Wang, Qingcai Jiang, Xuncheng Zhao, \\Rongfen Lin, Hong An\IEEEauthorrefmark{1}, Xiao Liang\IEEEauthorrefmark{1} and Lixin He\IEEEauthorrefmark{1} }
	\IEEEcompsocitemizethanks{
		\IEEEcompsocthanksitem M. Li, J. Chen, Q. Xiao, Q. Jiang, X. Zhao and H. An are with the University of Science and Technology of China. E-mail: \{mingfan, cjuns, xqbeida, jqc, zhaoxc\}@mail.ustc.edu.cn, han@ustc.edu.cn.
		
		\IEEEcompsocthanksitem F. Wang and R. Lin are with Tsinghua University. E-mail: \{f-wang20, lrf21\}@mails.tsinghua.edu.cn.
		\IEEEcompsocthanksitem X. Liang and L. He are with the CAS Key Lab of Quantum Information, University of Science and Technology of China, and  X. Liang is also with the Institute for Advanced Study, Tsinghua University. E-mail: lxxhlb@mail.ustc.edu.cn, helx@ustc.edu.cn.
	}
	\thanks{
		\IEEEauthorrefmark{2} Equal contributions.\\
		\IEEEauthorrefmark {1} Corresponding authors.\\
	}
}


\IEEEtitleabstractindextext{%
	\begin{abstract}
		Efficient numerical methods are promising tools for delivering unique insights into the fascinating properties of physics, such as the highly frustrated quantum many-body systems. However, the computational complexity of obtaining the wave functions for accurately describing the quantum states increases exponentially with respect to particle number. Here we present a novel convolutional neural network (CNN) for simulating the two-dimensional highly frustrated spin-$1/2$ $J_1-J_2$ Heisenberg model, meanwhile the simulation is performed at an extreme scale system with low cost and high scalability. By ingenious employment of transfer learning and CNN's translational invariance, we successfully investigate the quantum system with the lattice size up to $24\times24$, within 30 million cores of the new generation of sunway supercomputer. The final achievement demonstrates the effectiveness of CNN-based representation of quantum-state and brings the state-of-the-art record up to a brand-new level from both aspects of remarkable accuracy and unprecedented scales.
	\end{abstract}
	
	}

\vspace{1cm}
\maketitle


\IEEEdisplaynontitleabstractindextext

%
\IEEEpeerreviewmaketitle

\IEEEraisesectionheading{\section{Introduction}\label{sec:introduction}}

\IEEEPARstart{T}{he} quantum many-body problem, one of the most challenging research fields in modern physics, has attracted great attentions for revealing the fundamental natures of physical phenomena. However, the complexity for representing quantum many-body states grows exponentially with respect to particle number, which contributes it ranking among the most computationally intensive fields of science.

Compared to the exponential complexity of numerical algorithms\cite{kohn1999nobel,gygi2006large,hasegawa2011first}, the neural network offers a more effective way of improving system accuracy by increasing trainable parameters. Previous works on the fully connected network, such as RBM method\cite{carleo2017solving}, show great success in representing quantum many-body states. However, the redundant calculations from dense connections are not efficient, which cripple the representation ability for large-scale 2D quantum systems. Benefiting from local connectivity and parameter sharing, the CNN model owns more efficient computations and better ability for wave function representation. Moreover, the padding operation provides more supplementary boundary information, thus strengthening the ability for more difficult periodic boundary conditions.

Nowadays, the optimization method in deep learning, both theoretical and applied, is dominated by first-order gradient methods, like stochastic gradient descent\cite{you2017large, you2018imagenet,you2019large}. Yet despite these achievements, it is not powerful enough for optimizing quantum many-body systems. Specifically, the quantum system in ground state owns the global lowest energy, while the mainstream mini-batch gradient descent optimization may be easily trapped in some local minima. For the model with $N$ parameters, the first-order method theoretically captures the relation between $N$ parameters and loss function. But the second-order method makes it possible to further exploit the potential statistical correlations between (gradients of) different parameters by building the $N^2$ covariance matrix. In consequence, the second-order variational algorithm, like Stochastic Reconfiguration (SR) method\cite{neuscamman2012optimizing,sorella1998green}, can be principally interpreted as an effective imaginary time evolution in variational subspace.

So far, a CNN-based variational ansatz for variational quantum Monte Carlo (VQMC) method\cite{benedict2019quantum} has achieved competitive results\cite{liang2018solving,liang2021hybrid}. The ground state energy achieved on $10\times10$ square lattice is still far behind the state-of-the-art results\cite{nomura2021dirac}. Empirically, two major shortages impair the representation ability of the CNN-based algorithm: the limited number of network parameters (more generally variational parameters) and insufficient samples from Monte Carlo data sampling. To fully explore the potential ability of the CNN model, the above restrictions must be addressed, especially for attacking more challenging systems, which heavily depends on the modern parallel programming paradigm with high performance supercomputers.

To overcome the above issues, several types of optimization have been adopted on both process- and thread-level parallelism. First, the neural network computation is the most time-consuming step and involves no communication for intrinsically parallel Monte Carlo sampling. Through the high performance swDNN library, the hotspot operations are accelerated by offloading the hotspot functions onto many heterogeneous cores. Additionally, we transform the sequential sampling to a parallel version for mitigating the performance degradation from inadequate computations. Thus, one process maintains multiple independent chains and obtains multiple sampling states in one step, instead of collecting many states from one chain sequentially.

Considering the SR method requires much more samples to extract statistical information and construct the matrix, the program pattern significantly differs from mini-batch gradient descent. In this case, one optimization step needs millions of collected samples, which alleviates the pressure on data transmission. For another, the following optimization includes expensive and necessitated dense linear algebra calculations, where the matrix scale is consistent with the parameter number. On the sunway supercomputer, the challenge for increasing the parameters and scaling CNN model is resolved by distributed and parallel ScaLapack library\cite{choi1992scalapack}, which has assisted numerous HPC applications in solving large-scale linear equation systems.

Finally, transfer learning plays a vital role in confronting the exponential complexity of quantum systems for CNN's special translational invariance, which contributes CNN-based ansatz more scalable for large-scale quantum systems. From the view of the convolutional operation, the different size of spin lattices only influences the repeated times for sliding filters along the corresponding dimension, while the internal computation pattern for the local filter is independent. It is feasible that one CNN model can capture features from similar systems with different lattice size, because of the similarities in these systems. In general, transfer learning not only simplifies task difficulty and reduces the computational cost, but also opens up a promising route for confronting the challenge at unprecedented scales.

We evaluate our method on the spin-$1/2$ $J_1$-$J_2$ Heisenberg model on the square lattice, where quantum spin liquid can emerge in the intermediate region of $J2/J1\approx 0.5$, and the model is a candidate for high temperature superconductors. In the simulation of quantum many-body systems, larger lattice size leads to smaller boundary effects, which is crucial for the quantum system with long range correlations to capture the correct physics. By demonstrating our CNN-based solution, our results achieved remarkable progress over the state-of-the-art lattice size of  $18\times18$\cite{nomura2021dirac}. Overall, the quantum many-body systems on spin-$1/2$ $J_1-J_2$ model are firstly extended to the unprecedented scale for the entangled $24\times24$ square lattice, utilizing over 30 million heterogeneous cores on the new generation of sunway supercomputer, thus proving the CNN-based representation a promising tool for investigating challenges in quantum physics.

The contributions of this work are as follows:
\begin{itemize}
	\item We design and implement a low scaling and highly scalable CNN-based method for efficiently tackling the quantum many-body problem on new sunway supercomputer with modest parallel optimization towards the heterogeneous many-core processor;
	
	\item The CNN model with 106529 variational parameters is well optimized by millions of samples with the SR method and multiple stages of transfer learning, and eventually achieved the state-of-art accuracy for spin-$1/2$ frustrated Heisenberg model, especially firstly simulating the unprecedented scale of $24\times24$ quantum system;
	
	\item The great success of our CNN-based wave function representation introduces a promising paradigm for bridging the gap between HPC application and deep learning algorithm for efficiently overcoming the present difficulties, as well as providing a good reference for similar scenarios.
\end{itemize}

\setlength\extrarowheight{5pt}

\section{Background}
\subsection{Quantum system for spin-$1/2$ $J_1-J_2$ model}
\label{SR-background}
The properties of a quantum system are governed by its Hamiltonian. In nature, a quantum system is in ground state, and the energy is the lowest one of the Hamiltonian's eigenvalues, the lowest energy is the ground state energy.

Here we are investigating the two-dimensional spin-$1/2$ $J_1-J_2$ Heisenberg model, where the spin interactions are between the nearest-neighbor and next-nearest-neighbor sites. In our quantum simulations, the Heisenberg representation is applied. The operators are matrices and the states are vectors; the state of a spin can be either $|\uparrow\rangle$(up) or $|\downarrow\rangle$(down). The corresponding vectors are:

\begin{equation}
	|\uparrow\rangle=\begin{bmatrix}
		1 \\
		0
	\end{bmatrix},
	|\downarrow\rangle=\begin{bmatrix}
		0 \\
		1
	\end{bmatrix},
	\label{spin_vectors}
\end{equation}
The spin operators are defined by the Pauli matrices:
\begin{equation}
	\sigma^x=\begin{bmatrix}
		0 &1 \\
		1 &0
	\end{bmatrix},
	\sigma^y=\begin{bmatrix}
		0 &-i \\
		i &0
	\end{bmatrix},
	\sigma^z=\begin{bmatrix}
		1 &0 \\
		0 &-1
	\end{bmatrix},
	\label{pauli}
\end{equation}

It is straghtforward that both $|\uparrow\rangle$ and $|\downarrow\rangle$ are eigenvectors of operator $\sigma^z$. The states of $|\uparrow\rangle$ and $|\downarrow\rangle$ form a complete basis of the Hilbert space, that any vector can be represented by the linear combination of $|\uparrow\rangle$ and $|\downarrow\rangle$. Furthermore, the Pauli matrices with the identity matrix form a complete basis of a $2\times2$ matrix.

For a quantum system with $L^2$ spins, the basis can be written as the tensor product of each spin, thus the Hilbert space is $2^{L^2}$. Furthermore, the Hamiltonian is constructed by tensor products of Pauli matrices and the identity matrix, where the ground state is a linear combination of $2^{L^2}$ basis. The Hamiltonian $H$ reads:
\begin{equation}
	\label{J1-J2}
	H = \frac{J_1}{4}\sum\limits_{<i, j>}{}\sigma_i \cdot \sigma_j + \frac{J_2}{4}\sum\limits_{<<i,j>>}{}\sigma_i \cdot \sigma_j
\end{equation}
where $\sigma_i$ denotes Pauli operator, $\sigma_i\cdot\sigma_j=\sum_{k=x,y,z}\sigma_i^k\sigma_j^k$. $\langle i, j\rangle$ and $\langle \langle i,j\rangle\rangle$ denote interactions between the nearest-neighbor and next-nearest-neighbor sites, respectively.

Considering a 2D square lattice with the site number  $N=L\times L$, a basis is the configuration of all spins, denoted as: $|S\rangle=|s_1, s_2,\cdots, s_N\rangle$ where each spin value is $s_i=\pm 1$. The quantum state of a spin lattice can be represented in the form of $|\Psi\rangle=\sum_Sw(S)|S\rangle$, and $w(S)$ is the coefficient for the corresponding spin configuration $|S\rangle$. Once the $w(S)$ is obtained, the energy of the system can be calculated as:

\begin{equation}
	\label{sum-Es}
	E=\frac{\langle\Psi|H|\Psi\rangle}{\langle\Psi|\Psi\rangle}=\frac{1}{\sum_Sw(S)^2}\sum_Sw(S)^2\sum_{S'}\frac{w(S')}{w(S)}H_{S'S}
\end{equation}
where $H_{S'S}=\langle S|H|S'\rangle$.

In the variational quantum Monte Carlo (VQMC) method\cite{benedict2019quantum}, the energy $E$ and the gradients $G$ can be calculated via Monte Carlo sampling over spin configurations.

\begin{equation}
	\label{Es-Os-G}
	E=\langle E_s\rangle, G=\langle O_sE_s\rangle-\langle E_s\rangle\langle O_s\rangle
\end{equation}

where $E_s=\sum_{S'}\frac{w(S')}{w(S)}H_{S'S}$ , $O_s=\frac{1}{w(S)}\frac{\partial{w(S)}}{\partial{a_i}}$, and $a_i$ is the $i$-th variational parameter. The CNN model is then optimized by Stochastic Reconfiguration method. For every iteration and the collected samples, the final $\Delta$ for parameter update can be calculated:

\begin{equation}
	\label{delta}
	\Delta = \eta M^{-1}G
\end{equation}

where the $G$ is first order gradient of the iteration, and $M$ is a positive-definite covariance matrix,

\begin{equation}
	\label{convar-matrix}
	M_{kk^{'}}(p)=\langle O_k^{*}O_{k^{'}}\rangle-\langle O_k^{*}\rangle\langle  O_{k^{'}}\rangle
\end{equation}

$O_k(S)$ means the variational derivatives with respect to the $k$-th parameter. Note the covariance matrix $M$ may be non-invertible, thus we add a tiny shift for its diagonal elements to compute its Moore-Penrose pseudo-inverse.

\subsection{CNN-based wave function representation}
\label{CNN-background}
The major challenge of the quantum many-body problem is to represent the exponential amount of information in a wave function. However since the Hamiltonian contains local interactions and symmetries, it is possible to approximate the ground state in polynomial time. In our investigations, the approximation is done with CNN model, whose output can be regarded as the wave function coefficient.

\begin{equation}
	\label{wcnn}
	|\Psi_{CNN}\rangle=\sum\limits_{S}W_{CNN}(S)|S\rangle,
\end{equation}

For traditional regression problems, the output of CNN is usually continuously distributed for the continuous input. However, the sign generation of the quantum system\cite{voigt2000marshall} obeys discontinuous function with respect to flipping spin configurations. Taking the $J_1-J_2$ model as an example, when $J_2=0$,  the ground state exactly obeys the Marshall-Peierls sign rule (MPSR)\cite{szabo2020neural}, the sign of the wave function is $(-1)^{M_A}$, an obvious discontinuous, where $M_A$ is the magnetization of equivalent sublattice $A$. A possible solution for such conflict is to separate the wave function into two parts. The amplitude computation is completed by the CNN part and the sign generation is preconditioned as MPSR. In this case, the CNN wave-function is as follows:

\begin{equation}
	W(S)=(-1)^{M_A} W_{CNN}(S)
\end{equation}

Finally, the ground state many-body wave functions respect to the symmetry of the Hamiltonian. We could therefore also enforce the rotational symmetry manually,

\begin{equation}
	\label{sym-ws}
	\tilde{W}=\sum\limits_{i=0}^{3}W(\hat T^iS)
\end{equation}

and $\hat T$ is the rotation operator that rotates the spin configuration, the 2D square tensor, for 90 degrees.

\begin{figure}[tbp]
	\centerline{\includegraphics[scale=0.2]{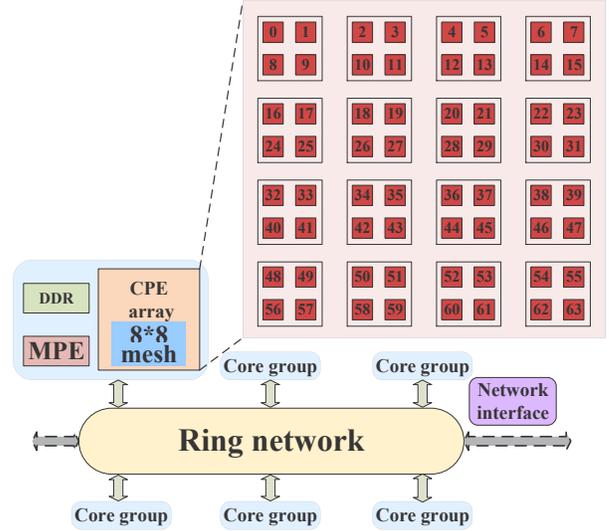}}
	\caption{Architecture of sw26010pro many-core processor.}
	\label{sw26010pro}
\end{figure}

\subsection{Sunway Processor Architecture}

The new sunway supercomputer is composed of the newly developed sunway many-core processors, and Fig. \ref{sw26010pro} presents the architecture of the latest heterogeneous sunway processor\cite{fu2016sunway,dongarra2016sunway}. The many-core processor (sw26010pro) consists of six core groups (CG), where six CGs are connected through a ring network. Each CG contains one management process element (MPE) and a cluster of 64 computing process elements (CPE) arranged in $8\times 8$ mesh.

Similar to the sw26010, MPE is deployed for handling management and communication with the ability of supporting interrupt functions, memory management. on the contrary, the CPEs are designed to achieve the maximum computing power for parallel programming. To obtain high performance on such heterogeneous cores, it is essential to follow these features and design appropriate parallel logic.

Differently, there are several convenient features of sw26010pro. First, the data transfer between CPEs is supported by the remote memory access mechanism, instead of previous register communication only between CPEs in the same row or column of the mesh. Besides, the local directive memory (LDM) of the upgraded CPE can be partly configured as a data cache, which significantly improves application performance. Especially for applications with regular memory access, data cache gains considerable performance speedup\cite{gao2021optimization}, even compared to the manual optimizations.

\begin{figure*}[tbp]
	\centerline{\includegraphics[scale=0.29]{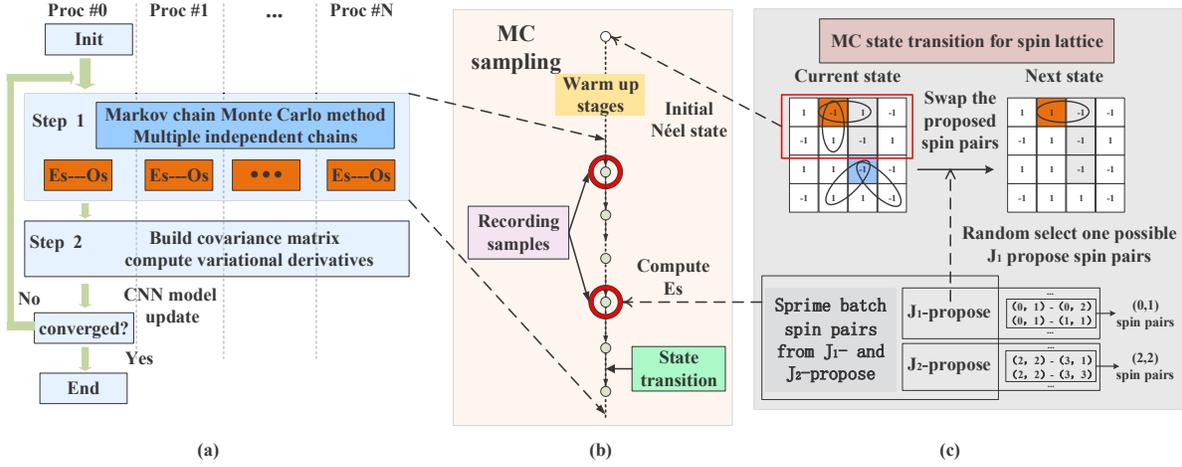}}
	\caption{Illustration of the basic framework. (a) MPI process parallelization of CNN variational algorithm. (b) Markov chain Monte Carlo sampling. (c) $J_1-J_2$ computation for Markov chain management.}
	\label{framework}
	\vspace{-0.3cm}
\end{figure*}

\begin{table}[h]
	\caption{Modification for scaling the CNN model}
	\resizebox{8.8cm}{!}{
		\begin{tabular}{|c|c|c|c|c|}
			\hline
			Description  & Layer type  & Small & Medium & Large \\ \hline
			\multirow{2}{*}{\begin{tabular}[c]{@{}c@{}}Building\\ Dimension\end{tabular}} & Convolution           & 6*6*3*3  & 16*16*3*3  & 32*32*3*3 \\ \cline{2-5}
			& TransposedConvolution & 6*6*1*3  & 16*16*1*3  & 32*32*1*3 \\ \hline
			\multicolumn{2}{|c|}{Toatal parameters for stacking blocks}                                                      & 3531     & 16249      & 106529    \\ \hline
		\end{tabular}
	}
	
	\label{building-block}
\end{table}

\section{Implementation and Optimization}

\subsection{Basic framework}
\label{bisic-framework}

For the CNN-based ansatz on spin-$1/2$ $J_1-J_2$ model, our previous papers\cite{liang2018solving,liang2021hybrid} demonstrate a comprehensive explanation from the perspective of physics. The network structure is entirely established by serially stacking the convolutional and maxpooling blocks. The model architecture is exactly same as prior literature\cite{liang2021hybrid} and this work simply increases the number of convolutional channels. Table \ref{building-block} presents the detailed modifications for scaling the network scale. Subsequently, another two models are adopted with 16249 and 106529 parameters over original 3531 parameters.

Fig. \ref{framework}a presents the basic framework for the CNN-based MC method. It consists of two indispensable parts: the MC entity of \textbf{Step 1} for collecting samples and the backend component of \textbf{Step 2} for data handling. The computation starts from an initial state with randomly initialized network parameters. The "Initialization" on process \#0 includes a necessary warmup stage for preparing the spin lattice. It performs repeatedly state transitions until the CNN output becomes stable. Note the enormous search space, the initial spin configuration usually begins at the N\'{e}el state, where any two neighbor spins point in opposite directions, marked red in Fig. \ref{framework}c. After process \#0 broadcasts the initialized parameters and warmup spin lattice, the main loop starts MC sampling, which repeatedly performs forward and backward computation for gathering $(E_s, O_s)$ in Eq.\ref{Es-Os-G}. The latter framework then computes the $\Delta$ for model update. The iteration of the main loop continues until the energy approaches to a given criterion.

Fig. \ref{framework}b describes the step for Markov chain Monte Carlo (MCMC) method\cite{geyer1992practical}. It consists of both routines: the state transition of  Markov chain and the Monte Carlo sampling. Given the spin configuration $S$, the probability amplitude is calculated by the CNN model. Following the Eq. \ref{wcnn}, $W_{CNN}(S)$ controls the state transition for the Markov chain. Note that the output of CNN ignores sign generation problem, we thus choose $P=(\frac{W_{CNN}(S_{next})}{W_{CNN}(S)})^2$ as final standard in determining state transition. To avoid model overfitting in extreme conditions and expand the search space, the random walk criteria is decided by a random probability $P \geq r$, where $r\in(0,1)$. In general, we should explore as much broader space so that it captures the physical characteristic to represent the whole Hilbert space, instead of acquiring several extreme values for limited local space.

Fig. \ref{framework}c presents an example $4*4$ spin lattice for explaining the $J_1-J_2$ model, in Eq \ref{J1-J2}. When the nearest-neighbor (next-nearest-neighbor) spins in opposite direction, there comes the $J_1$ ($J_2$) interaction. Considering the propose batch iterates every spin point of the lattice, we only compute half of the four bonds (e.g. the right and below neighbors for $J_1$ interaction). The spin pairs are recorded as the $J_1$- and $J_2$-propose batch. The example spin pairs of $(0,1)$ for $J_1$ interaction and $(2,2)$ for $J_2$ interaction are listed, whereas the position $(0,1)$ contains no $J_2$ interaction. The state transition is randomly selected from the candidate $J_1$-propose batch, while the sampling is computed over sprime\_batch for both interactions. The final sum is calculated by the whole $W(S^{'})$ to compute the energy, following in Eq. \ref{sum-Es}.

Another argument, sampling gap, manages the difference of selected samples and eventually influences the computation of $\Delta$. To ensure the independence between the samples, we try $L^2$ times of spin flip in the Markov chain, without calculating $E_s$ and $O_s$.

With tremendous computation for sprime\_batch and sampling gap, the amount of forward is roughly $3L^2$ times that of backward ($2L^2$ for $E_s$ and $L^2$ for sampling gap only). Moreover, the increasing number of parameters requires much more samples to extract the statistical correlations as defined by the SR method. For the model with 106529 parameters on $24\times24$ system, one optimization step needs billions of CNN forward execution, which accounts for the main part of the execution time.

As for the communication pattern, the independent Markov chains for importance sampling on different processes are intrinsically parallel, which involves little communication and can be directly distributed across all participating workers. Theoretically, the communication cost for computing the average gradients grows along with the parameter number, while the second-order method includes expensive dense linear algebra calculations. The covariance matrix scale is $N^2$, where $N$ is the number of variational parameters. It is not influenced by the system scale or the number of samples. For instance, the spin lattice of $6\times6$ system shares the same number of parameters and corresponding matrix as that of $24\times24$ system, which further mitigates communication pressure over the computational complexity.

\subsection{Performance Optimization}
\label{parallelization}

With the naive framework, modest optimization techniques have been considered for confronting the immerse computational cost and the tricky computation for the SR method along with the increasing parameters. Based on original inter-process parallelization for the independent Markov chains, the sequential MC sampling is further optimized. It increases the parallel granularity and eliminates the serial dependency, as well as better adaptation to modern heterogeneous devices. Meanwhile, the severe restriction to large-scale matrix computation is solved by parallel and distributed ScaLapack library.

\subsubsection{Parallel optimization for sequential sampling}

In our previous works, the program is executed on several distributed homogeneous nodes for tens of processes with direct MPI parallel style. Each process collects thousands of $\langle E_s, O_s\rangle$ samples. However, such implementation meets limitations when it comes to modern heterogeneous accelerators, like Google TPU, NVIDIA GPU, and sunway heterogeneous processor. It is impossible to take full advantage of such powerful computing resources within single Markov chain. The execution for the CNN model with a small batch size suffers significant performance degradation.

\begin{figure}[tbp]
	\centerline{\includegraphics[scale=0.35]{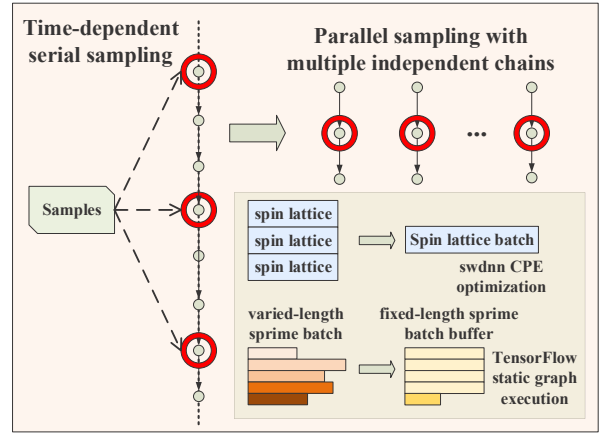}}
	\caption{MC sampling parallelization inside single process.}
	\label{parallel-mc}
	\vspace{-0.3cm}
\end{figure}

We expand the intrinsically parallel importance sampling not only among the MPI processes for distributed nodes, but also inside the single process. This modification replaces the sequential MC sampling by multiple independent Markov chains as illustrated in Fig. \ref{parallel-mc}. With multiple independent chains, the state transition for one step contains multiple parallel CNN forward computations. Naturally, these input spin lattices are combined into a batch, and computed by one time of the CNN forward operation for the respective results.

On the new generation of sunway supercomputer, the actual computation is finally accelerated by the 64 CPEs of the sw26010pro processor. Based on our previous study about computation parallelization on CPE cluster\cite{li2021swflow}, the convolution operation provides better performance with the batch size being times of 8. The parallel design for increasing batch size assists a lot in improving the system performance. Moreover, the sprime\_batch from different spin lattice often owns varied lengths of $J_1$ and $J_2$ propose batch, and the computation of $Es$ needs to sum all $W(S^{'})$ of the propose batch, in Eq. \ref{sum-Es}. In our parallel implementation, multiple repeated CNN forward with varying lengths can be replaced by the fixed-length execution. With the same batch size, the repeated forward executions may reuse the existing tensor resource and improve the performance, especially for the static graph execution of TensorFlow.

To realize the parallel implementation, the serial algorithm needs necessary modification to maintain multiple independent chains. There may exist sharp areas where the probability is completely wrong and makes the chain trapped in extreme conditions. The parallel MC tasks on the sunway supercomputer eventually deploy millions of independent Markov chains, which inevitably incurs more danger of abnormal states. A worse condition is the wrong $E_s$ from an incorrect chain is far wary from the standard result, which contaminates all other samples. In this case, an additional check procedure is appended after the sampling of $E_s$, which is judged by certain physics rules. To ensure the robustness of the simulation, we terminate the problematic Markov chain and start a new one.

Overall, our implementation reduces the gap between inadequate computation and extreme computing power. Instead of realizing such sequence execution inside TensorFlow framework, the harsh dependence issues are circumvented from high-level application characteristics. Specifically, the state transition is determined by the random probability, which theoretically proves the parallel sampling and avoids sequential influences. The final experiments also demonstrate the effectiveness of such a radical design for the spin-$1/2$ model. Except for certain problems with strong limitations\cite{kim2008chiral}, it can be referred for similar MC problems.

\subsubsection{Parallel computation of $\Delta$}
After sufficient samples of $E_s$ and $O_s$ are collected from the MC entity, the backend component is responsible for computing $\Delta$ based on the SR method. As introduced in Sec. \ref{SR-background}, the algorithm consists of several steps:

\begin{itemize}
	\item 1) computing the average result, $\langle E_s\rangle $, $\langle O_s \rangle$  and the first order gradient $G$, from the collected $E_s$ and $O_s$;
	\item 2) constructing the covariance matrix as Eq. \ref{convar-matrix}, $M={O}^{T}_{mn}O_{mn}-\langle O_s \rangle^{T}\langle O_s\rangle+shift \times I_{nn}$, where $O_{mn}$ means the matrix formed by total $m$ collected $O_s$ data and each $O_s$ contains $n$ variational parameters, $shift$ for keeping the matrix invertible;
	\item 3) solving the linear system to compute $\Delta=\eta M^{-1}G$ as Eq. \ref{delta}, with $\eta$ represents the learning rate, and applying the model update.
\end{itemize}

\begin{figure}[tbp]
	\centerline{\includegraphics[scale=0.3]{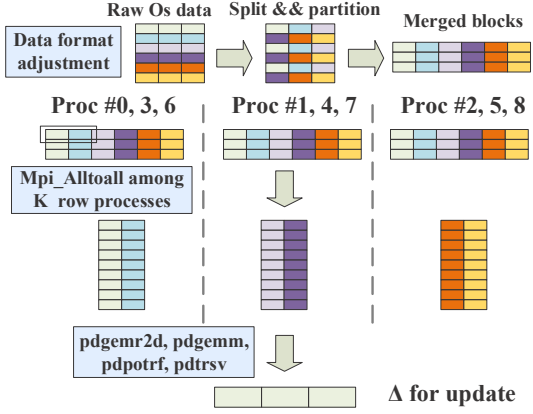}}
	\caption{The example 2D-mesh distributed ScaLapack computation for 9 processes with 6 as single batch size.}
	\label{scalapack}
\end{figure}

The SR optimization method is mostly applied for many QMC problems. The most challenging parts are to construct the covariance matrix and to solve the linear systems for a large number of variational parameters. It essentially involves considerable linear algebra computation, which is more related to the underlying hardware conditions than the high-level applications. For example, the corresponding adaptation is the Intel Math Kernel Library (MKL) on the Tianhe supercomputer, while the project for NVIDIA multi-GPUs environment is implemented with "cublasXt" and "MAGMA" library. On the sunway supercomputer, the low-level ScaLapack library provides better support for parallel distributed large-scale linear systems, which becomes the preferred solution for such HPC application scenarios.

Considering the detailed communication inside ScaLapack library is transparent, we briefly introduce the upper program pattern for handling the evenly distributed samples among all participated workers. For the instance of 9 processes arranged in 3*3 mesh, the single batch size is 6, i.e., each process owns 6 samples of $E_s$ and $O_s$. Fig. \ref{scalapack} presents the procedures for data operations.

At first, it adjusts the data format of the original $O_{s}$ matrix, where each $Os$ is divided into 3 blocks (the number is equal to the row process number). The total 18 blocks of 6 $O_s$ are merged into a new matrix with 3 rows. Secondly, the merged blocks is then arranged into the 2D-mesh distributed matrix by the MPI\_Alltoall invocation for a group of processes in the same row. Different processes in the same column perform the same operation on their corresponding samples, like the process \#0, process \#3 and process \#6. The local communication in the same row happens among the 2D meshing processes and reduces global network traffic. Finally, the ScaLapack subroutines, like 'pdgemm', 'pdpotrf', can be directly called to build the distributed covariance matrix. With the intermediate result of 'pdpotrf', a more straightforward solution is to compute the $\Delta$ with twice invocations of 'pdtrsv', instead of the matrix inverse.

\begin{figure}[tbp]
	\centerline{\includegraphics[scale=0.25]{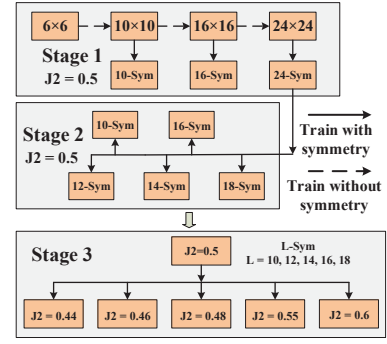}}
	\caption{Three stages application of transfer learning.}
	\label{Transfer-learning}
	\vspace{-0.5cm}
\end{figure}

\subsection{Transfer learning}
\label{trans-train}
Transfer learning is initially proposed for addressing the training of neural networks and has achieved great success in many application scenarios. The core idea is to store knowledge acquired for solving one problem and apply it to related ones. It principally overcomes the isolated learning paradigm and bridges the gap between different problems by practical ties. From the practical standpoint, transferring information between tasks has superior advantages of improving the overall efficiency and expanding the present approach towards more practical or complicated problems.

Fig. \ref{Transfer-learning} presents three stages for applying transfer learning in our work. The first stage describes that it is feasible that utilizing the same model captures the corresponding feature for different scales of the input square lattice. For the $6\times6$ system, model optimization starts at the randomly initialized weight and can be trained to a stable state with cheap computational cost. We then optimize the large-scale system with the pretrained model. By four stages of transfer learning, we finally obtain a well-optimized state for the $24\times24$ system. After the last fine-tuning stage with the lattice symmetry, the model eventually conquers the challenge for $24\times24$ system for spin-$1/2$ on the $J_1-J_2$ model.

At first, the rotational symmetry of the 2D square lattice is one of the inherent nature for quantum systems. Enhancing such a feature means more accurate results with extra computational cost, where it requires 4 times computation cost from Eq. \ref{sym-ws}. In the early stage of training, the model is far from the ideal state and the optimization requires rough direction, rather than the accurate value. Thus, the symmetry is initially ignored at the early stage, and can be manually supplied for the last fine-tuning stage.

Secondly, the property of translational invariance fits the quantum system on the square lattice. The internal computations of the convolution kernel are not related to the size of input data. The input spin lattice with different size influences the times to move the convolution kernel, while it keeps the same computation pattern. For the $J_1$-$J_2$ model under periodic boundary condition, shifting the lattice does not change the property of the quantum system. Thus the model optimized on the $6\times 6$ lattice can be used as an initial network on the $10\times10$ lattice, which significantly reduces the optimization time for large lattices.

In principle, the final optimized model for $24\times24$ system can be regarded as the most powerful representation of wave function for spin-$1/2$ $J_1-J_2$ model. Empirically, the well-optimized model for $24\times24$ system captures the characteristics for similar systems. Therefore, two subsequent stages are conducted for thoroughly evaluating transfer learning on quantum systems. For \textbf{Stage 2}, other systems with different scales are evaluated with the same $J_2/J_1$ value. For \textbf{Stage 3}, further development of transfer learning on more challenging problems where the value of $J_2/J_1$ is flexible, ranging from 0.44 to 0.6, which generally saves enormous computational cost for optimization from scratch. 

\begin{table}[t]
	\caption{Experimental parameters.}
	\resizebox{8.8cm}{!}{
		\begin{tabular}{|c|c|c|}
			\hline
			Description                                  & Parameter                                   & Remark or value                                                  \\ \hline
			\multirow{3}{*}{Application}     & CNN model (parameter number)          & 16249,106529                                                     \\ \cline{2-3}
			& Quantum system (2D spin-lattice size) & $6\times6$, $10\times10$, $16\times16$, $24\times24$                                            \\ \cline{2-3}
			& Transfer learning on system condition       & Scale, Symmetry, J2 interaction \\ \hline
			\multirow{3}{*}{\begin{tabular}[c]{@{}c@{}}System\\environment\end{tabular}}
			& Overall software stack                           & TensorFlow, swdnn, MPI, ScaLapack                                \\ \cline{2-3}
			& Maximum number of CGs used                  & 490000 (one process maps one CG)                                 \\ \cline{2-3}
			& Maximum number of cores used                & 31850000                                 \\ \hline
		\end{tabular}
	}

	\label{envs}
	\vspace{-0.5cm}
\end{table}

\begin{figure}[bp]
	\vspace{-0.5cm}
	\centerline{\includegraphics[scale=0.17]{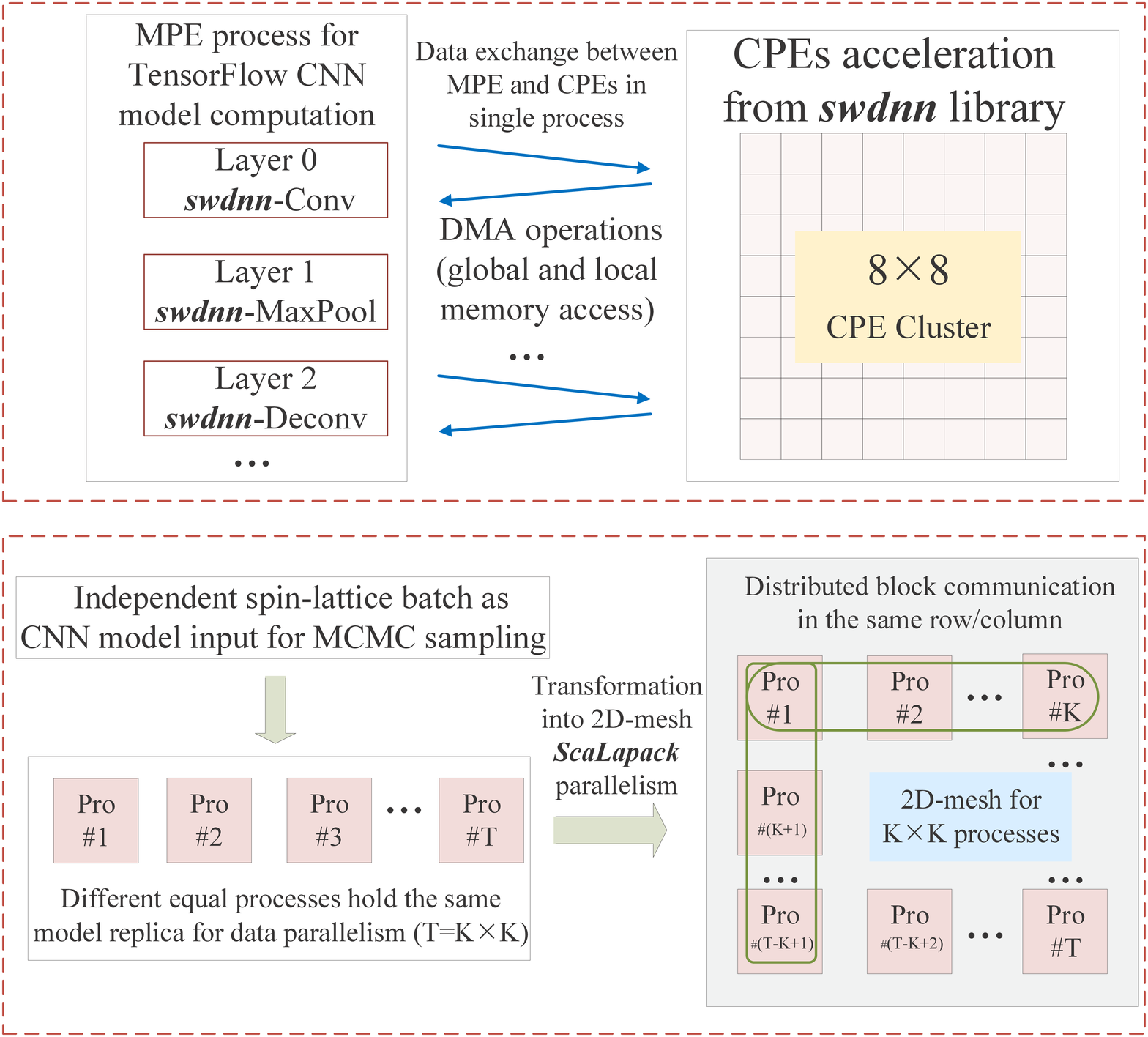}}
	\caption{Flowchart of swdnn and ScaLapck: intra- and inter- process implementation on new sunway supercomputer.}
	\label{swsystem}
	\vspace{-0.5cm}
\end{figure}

\section{Experiments and Evaluation}
We demonstrate our CNN-based method by solving the spin-$1/2$ $J_1-J_2$ model on square lattice for the periodic boundary conditions. The tests were done on the new generation sunway supercomputer, and the specific experimental parameters are provided in Table \ref{envs}.

To obtain the ground state energy, all processes execute two steps repeatedly in Fig. \ref{framework}a, where the flowchart of swdnn and ScaLapack on new sunway supercomputer is presented in Fig. \ref{swsystem}. Every process locates inside each CG, where a cluster of 64 CPEs assists the MPE for hotspot operations. Through the swdnn library\cite{fang2017swdnn}, the corresponding model computation on MPE is divided into multiple computing layers on the CPE cluster, where the data accessed by rules are transferred between the main memory and local memory through DMA operations. The MC sampling processes execute in data parallel style, where different processes hold the same model replica and handle independent input data. With the additional data redistribution in Fig. \ref{scalapack}, these equal workers are organized into 2D-meshing processes. The global communication is then adapted into local style (group processes in the same row or column) by distributed ScaLapack\cite{choi1992scalapack}. In this case, three aspects of experiments and performance analysis are elaborated.

For the single process, the speedup from heterogeneous CPE cluster for CNN-based MC sampling is evaluated on the hotspot operations. For the scalability and distributed environment, we evaluate the performance of ScaLapack and the parallel efficiency of the MC part. For the small model with 16249 parameters, the quantum system size is scaled to $16\times16$ system; another large model with 106529 parameters is well optimized to confront the $24\times24$ system. This paper merely presents the ground state energies along with the model optimizing, especially for examining the effectiveness of transfer learning on quantum many-body problems. More systematic results and discussion on the physical properties, including the comparison with state-of-the-art results, will be presented in a separate paper\cite{unpublished}.

\begin{figure}[tbp]
	\centerline{\includegraphics[scale=0.5]{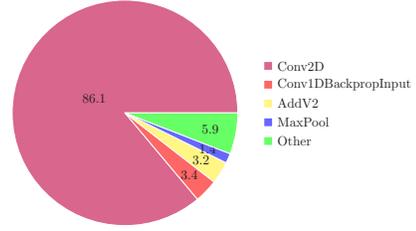}}
	\caption{Hotspot operations of CNN forward computation.}
	\label{CNN-computation-pie-1}
	\vspace{-0.5cm}
\end{figure}

\begin{figure}[bp]
	\vspace{-0.5cm}
	\centerline{\includegraphics[scale=0.66]{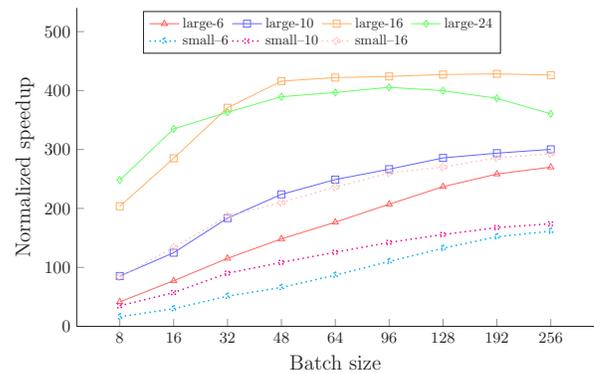}}
	\caption{Convolution operation performance speedup by CPE cluster: the small model (16249 parameters) is evaluated for systems scale as $6\times6$, $10\times10$ and $16\times16$; the large model (106529 parameters) is further evaluated for $24\times24$ system.}
	\label{conv-op}
\end{figure}

\subsection{Single process performance}

As introduced in Sec. \ref{bisic-framework}, the CNN forward computation is the most time-consuming part for computing millions of samples, especially for a large system. We thus test the execution time of the CNN forward computation by the MPE version. The statistics of operations time ratios for $10\times10$ square lattice for 106529-parameter model are provided in Fig. \ref{CNN-computation-pie-1}. The $Conv2D$ operation is the major hotspot and four operations account for over 90\% of total execution time.

\begin{table*}[t]
	\centering
	\caption{Scalability results for the computation of ScaLapack library and CNN execution.}
	\resizebox{16cm}{!}{
		
		\begin{tabular}{|c|c|c|c|c|c|c|c|c|c|c|}
			\hline
			\multirow{2}{*}{model scale} & \multirow{2}{*}{Process} & \multirow{2}{*}{ScaLapack(s)} & \multicolumn{4}{c|}{MC-CNN(s)} & \multicolumn{4}{c|}{Parallel Overhead(\%)} \\ \cline{4-11} & & & 10*10-nosym & 10*10-sym & 16*16-nosym & 16*16-sym & 10*10-nosym & 10*10-sym & 16*16-nosym & 16*16-sym \\ \hline
			\multirow{4}{*}{\begin{tabular}[c]{@{}l@{}}16249\\ parameters\end{tabular}}
			& 256   & 6.43  & 14.2  & 35.1  & 41.3  & 139.7 & 0.453  & 0.183  & 0.155  & 0.046  \\
			&1024  & 3.87  & 12.2  & 31.9  & 38.3  & 136.8 & 0.317  & 0.121  & 0.101  & 0.028  \\
			&4900  & 3.81  & 11.6  & 32.7  & 38.5  & 137.2 & 0.328  & 0.117  & 0.099  & 0.027  \\
			&10000 & 4.15  & 12.5  & 34.2  & 38.3  & 137.1 & 0.332  & 0.121  & 0.108  & 0.03   \\ \hline
			\multirow{4}{*}{\begin{tabular}[c]{@{}l@{}}106529\\ parameters\end{tabular}}
			&1024  &  136.46	 &  171.7  &  277.3  &	296.1  &  797.4  & 	0.795  &  0.492  &  0.461  &  0.171  \\
			&4900  &  82.77   &  119.2  &  215.5  &  253.2  &  754.3  &  0.694  &  0.384  &  0.327  &  0.109  \\
			&10000 &  68.46   &  104.4  &  208.7  &  239.4  &  742.1  &  0.656  &  0.328  &  0.286  &  0.092  \\
			&40000 &  55.17   &  95     &  197.2  &  226.9  &  727.7	 &  0.581  &  0.279  &  0.243  &  0.076  \\ \hline
		\end{tabular}
		
	}
	
	\label{small-scal-table}
\end{table*}

\begin{figure*}[tbp]
	\centering
	\begin{subfigure}[b]{.45\textwidth}
		\includegraphics[width=\textwidth]{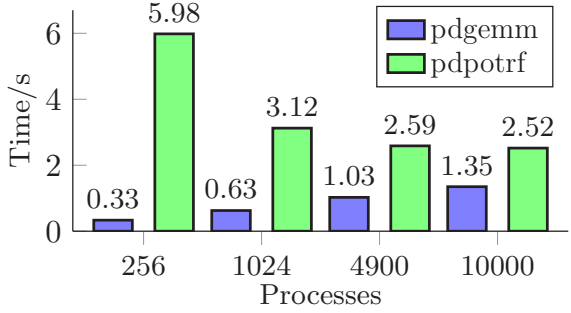}
		\caption{Small model with 16249 parameters.}\label{fig:gull}
	\end{subfigure}
	\begin{subfigure}[b]{.45\textwidth}
		\includegraphics[width=\textwidth]{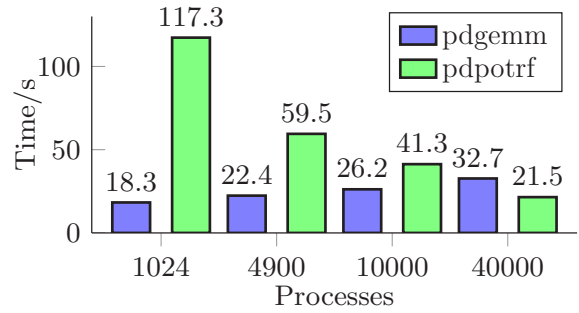}
		\caption{Large model with 106529 parameters.}\label{fig:tiger}
	\end{subfigure}
	\caption{Scalability results for the detailed ScaLapack hotspot functions: the computational cost of pdgemm is related to the number of samples; while the pdpotrf is independent from quantum system scale and relevant to model parameters.}
	\label{small-scalability}
	\vspace{-0.5cm}
\end{figure*}

Fig. \ref{conv-op} describes the normalized speedup of swDNN for the performance of convolution operation with the change of batch size. It is clear that the increasing amount of computation improves system performance, from three aspects of the increase for CNN model, spin lattice and batch size. For the small model, the performance grows clearly along with the overall batch size and system scale. The larger amount of computation contributes to better performance until performance saturation. For the large model, swDNN finally achieves considerable performance improvement by 300x speedup  (up to 400x for the $16*16$ and $24*24$ tasks).

Compared to traditional computer vision tasks, the computation of convolution operation for our task is far less. For one thing, the image format from the Imagenet dataset can be large as $224\times224\times3$. On the other hand, the channel number is often larger than 256. To this end, the limited computation of convolution operation is hard to make full use of the processor's computing ability. 

The other three types of operation present similar behavior. Moreover, their internal calculation is far less than the $Conv2D$ operation and can be regarded as memory-intensive computation. Compared to the MPE version, swDNN library finally gains 12x, 21x, 24x speedup for $Maxpool$, $Conv1DBackInput$ and $Add$, respectively.

\subsection{Scalability}
In the parallel scalability tests, we record the total execution time of a complete iteration, which includes the intrinsically parallel MC time and the elapsed time for SR computation from ScaLapack library. To evaluate the parallel overhead of the ScaLapack library, two models with different parameter number are investigated with the increasing number of processes, where we additionally record the execution time of the hotspot functions, 'pdgemm' and 'pdpotrf'. The process number for the small model is scaled up to 10000; while the corresponding processes number for the large model varies from 1024 to 40000. The quantum system size for the MC part is $10\times10$ and $16\times16$, where we additionally test the CNN forward with the symmetry not considered.

In principle, the MC sampling only performs CNN computation, which is influenced by the CNN computation and independent to the number of processes. On the contrary, the ScaLapack computation is relevant to the parameter number and influenced by the process number, but not related to the system scale. The overall scalability results are presented in Table \ref{small-scal-table}. The detailed results of the ScaLapack time ('pdgemm', and 'pdpotrf') is shown in Figure \ref{small-scalability}.

\begin{figure}[t]
	\centerline{\includegraphics[scale=0.6]{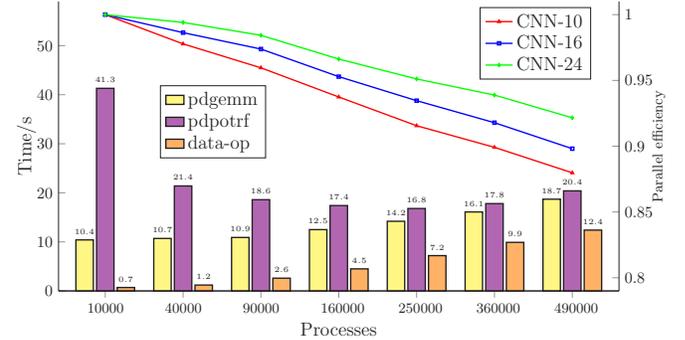}}
	\caption{The performance evaluation for large-scale system: the hotspot details of ScaLapack operation and the CNN execution evaluated on three types of quantum system scale.}
	\label{large-scalability}
	\vspace{-0.5cm}
\end{figure}

\begin{figure*}[thbp]
	\centering
	\begin{subfigure}[b]{.3\textwidth}
		\includegraphics[width=\textwidth]{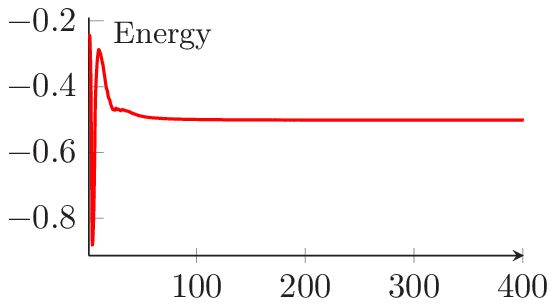}
		\caption{$6\times6$ system without symmetry.}\label{fig:mouse}
	\end{subfigure}
	\begin{subfigure}[b]{.3\textwidth}
		\includegraphics[width=\textwidth]{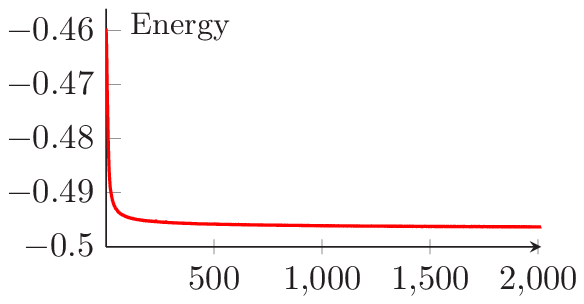}
		\caption{$10\times10$ system without symmetry.}\label{fig:gull}
	\end{subfigure}
	\begin{subfigure}[b]{.3\textwidth}
		\includegraphics[width=\textwidth]{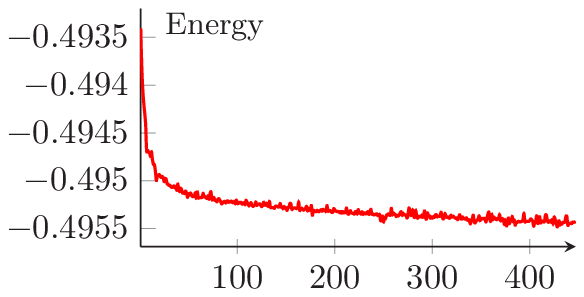}
		\caption{$16\times16$ system without symmetry.}\label{fig:tiger}
	\end{subfigure}
	\vspace{0.3cm}
	
	\begin{subfigure}[b]{.37\textwidth}
		\includegraphics[width=\textwidth]{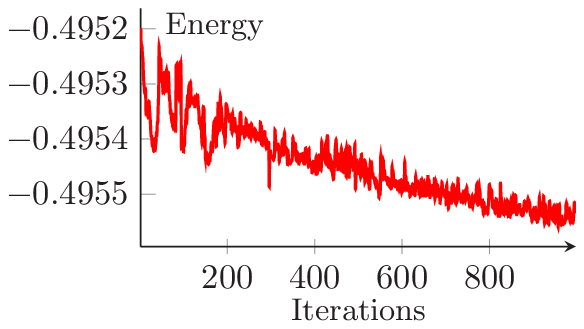}
		\caption{$24\times24$ system without symmetry.}\label{fig:gull}
	\end{subfigure}
	\begin{subfigure}[b]{.37\textwidth}
		\includegraphics[width=\textwidth]{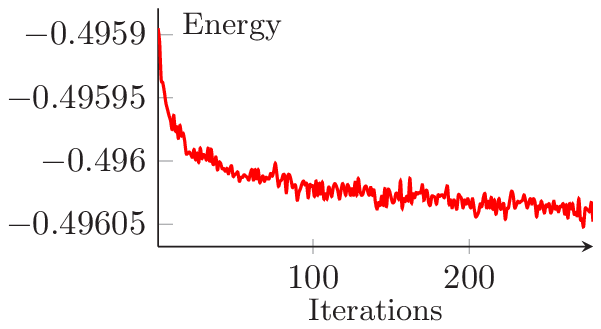}
		\caption{Final stage with symmetry on $24\times24$ system.}\label{fig:tiger}
	\end{subfigure}	
	\caption{Five stages of transfer learning for optimizing large model (106529 parameters) on $24\times24$ quantum system.}
	\label{main-train-24}
	\vspace{-0.5cm}
\end{figure*}

For the small model, the increasing of parallel processes improves the SR computation efficiency with the distributed ScaLapack, from 6.43 seconds to 4.15 seconds. However, the parallel overhead of 10000 processes is slightly higher than that of the 4900 processes. With fixed matrix scale as $16249*16249$, the increasing of MPI processes introduces more computation overhead of "pdgemm", while the parallel optimization of "pdpotrf" may saturate with the process number increasing, because of the small distributed blocks. Overall, the parallel overhead reaches 12.1\% and 3\% for the $10\times10$ and $16\times16$ system on 10000 processes for computing $\Delta$ with 160000 MC samples. With the inherent symmetry ignored, the corresponding overhead grows to 33.2\% and 10.8\%, owing to the decreased proportion for the CNN computation. For the large model, the "pdpotrf" time decreases from 117.27 seconds to 21.45 seconds with the increasing parallel processes, and the overall ScaLapack performance gains 2.47x speedup. The parallel overhead comes to 27.9\% and 7.6\% for $10\times10$ and $16\times16$ system on 40000 processes for computing $\Delta$ with 640000 samples.

Further evaluation for the large-scale system is delivered for 106529 parameters. The process number varies from 10000 to 490000 for three different system scales. The ScaLapack time and CNN parallel efficiency results are drawn in Fig. \ref{large-scalability}. Similar to the small model, the large model reaches the saturation point on 90000 processes. Except for the original hotspot functions of "pdgemm" and "pdpotrf", other data operations also cripple the performance, like the data format adjustment, "pdgemr2d", and "pdtrsv". For the 10000 processes, the elapsed time is negligible; while it grows to 12.4 seconds for 490000 processes.

Overall, the parallel efficiency reaches 86.7\% for $10\times10$ (89.4\% for $16\times16$) quantum systems on 490000 processes for computing $\Delta$ with nearly 4 million MC samples. Benefiting from the same ScaLapack time and the increasing proportion for the CNN computation, the $24\times24$ system scales from 650000 cores to over 30 million cores with a parallel efficiency of 92.5\% for a 49-fold increase of the MC samples.

Finally, we note that in our application, the first sampling stage is based on the Monte Carlo method and adopts the data parallelism style for CNN computation as shown in Fig. 6, which is intrinsically parallel and involves little communication. For the later SR optimization, the communication pressure for 2D-mesh ScaLapack grows as the root of the total number of processes. Furthermore, the communication overhead is theoretically not influenced by the quantum system scale, and therefore may achieve better scaling efficiency for large-scale quantum systems. We therefore believe the application can easily be scaled up to the whole new generation of sunway supercomputer.

\subsection{Transfer learning}

This section illustrates the effect of transfer learning, especially for the large-scale quantum systems. Firstly, transfer learning on scaling the system scale is examined by multiple rounds from $6\times6$ system to $24\times24$ system. Then, several comparison experiments are introduced for evaluating the connection between model representation ability and parameter number. In addition, further evaluations are proposed on validating the performance for various system scales and different $J_2/J_1$ values for the $J_1-J_2$ model.

\subsubsection{Five stages of optimization for $24\times24$ system}
As introduced before, the main optimization can be divided into 5 stages for solving the $24\times24$ system. The first stage is to train the CNN model from randomly initialized parameters on the $6\times6$ system. The initial steps are so unstable that we remove the warmup stage and increase the sampling frequency with a small sampling gap. Markov chain starts from the N\'{e}el state and keeps minor differences between two successive samples for alleviating the complex quantum interactions. Once steadily passing the initial stage and the energy becomes stable, we are able to increase the sampling gap ($L^2$ for the $L\times L$ system), for both purposes of avoiding local minima and accelerating optimization.

With the well-optimized model from $6\times6$ system, the model can be directly used for optimizing the $10\times10$ system after a warmup stage for Markov chain initialization. The CNN model goes through another round of $16\times16$ optimization for final $24\times24$ system. After the fine-tuning stage with the rotational symmetry, our CNN model successfully masters the wave function for the spin-$1/2$ $J_1-J_2$ model on $24\times 24$ square lattice, with $J_2/J_1$ is 0.5.

The overall training curves for our five-stage optimization are drawn in Fig. \ref{main-train-24}. The initial stage of $6\times6$ training starts from N\'{e}el state, where the spin lattice owns all $J_1$ interaction and no  $J_2$ interaction. The corresponding energy is around -0.25. After the violent fluctuations on the initial several steps, the training results present a steady curve for optimization towards lower energy.

For the $6\times6$ system without symmetry, the final energy is optimized to -0.502 in 420 training steps. Through transfer learning, the initial energy for the $10\times10$ system is around -0.46 and finally achieves -0.496 with 2000 steps. From the view of physics, scaling the system from $10\times10$ to $16\times16$ is more stable and reasonable. For this reason, the initial state for $16\times16$ system starts at the energy around -0.493, and ends around -0.4955 with 400 iterations. Subsequently, the initial energy of the $24\times24$ system is -0.4952, and the model state ends at the energy of -0.4955. Meanwhile, the more and more accurate energy from five states reveals the effectiveness of our whole optimization with the gradual training progress.

The last stage for complementing the system with rotational symmetry starts at -0.4959. Within only 200 steps of optimization, the system energy obtained from the CNN model is over -0.49605. Compared to the state-of-the-art $18\times18$ system, the firstly solved $24\times24$ system with the excellent ground state energy proves the effect of transfer learning and demonstrates the great success of the CNN-based wave function representation. The state-of-art accuracy result pushes the deep learning method into practical application over theoretical research.

\subsubsection{Comparison between different models}
We further conduct experiments on optimizing the $10\times10$ and $16\times16$ system to study the connection between the model representation ability and the parameter number. The overall optimization stages are arranged in accordance with \textbf{Stage 1} in Sec. \ref{trans-train}, without the last stage for the $24\times24$ system. The experiments are evaluated for the models with 16249 and 106529 parameters, respectively.

\begin{figure}[tbp]
	\centering
	
	\includegraphics[height=6.3cm]{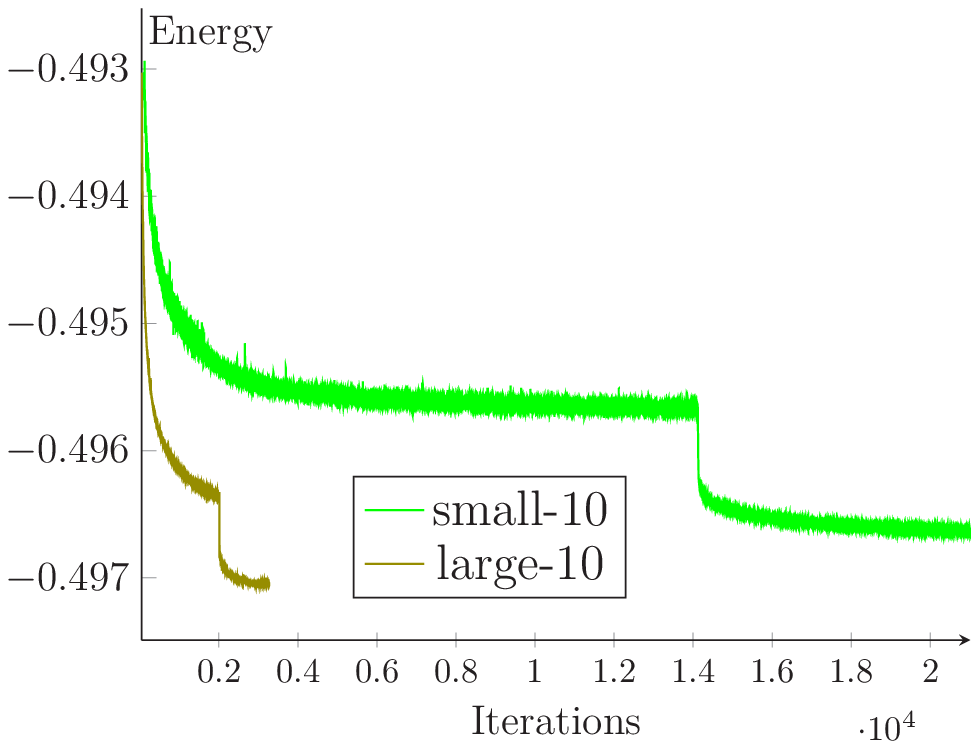}
	\llap{\raisebox{3cm}{
			\includegraphics[height=3.2cm]{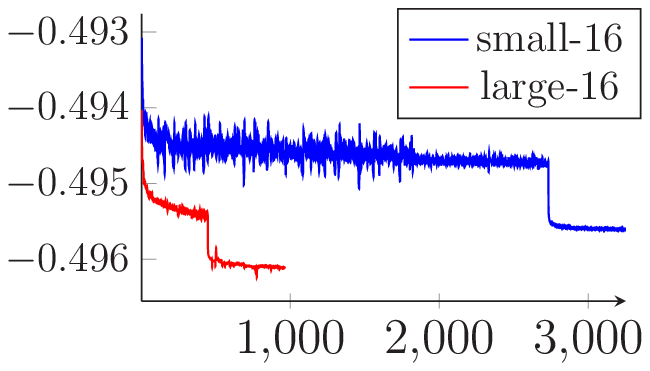}
		}
	}
	\caption{Comparison on $10\times10$ and $16\times16$ quantum systems with different model scale: the small model with 16249 parameters and the large model with 106529 parameters.}
	\label{more-parameters}
	\vspace{-0.3cm}
\end{figure}

Fig. \ref{more-parameters} presents our training curves with an evident observation. The amount of variational parameters, for the CNN-based variational ansatz, plays an essential role in representing the wave functions. For the $10\times10$ system, the small model with over 20000 optimization steps gains the energy result as -0.49665 ($10^{-3}$ relative error defined by $(E-E_{best})/E_{best}$). It is a great improvement over the baseline energy, -0.4955 ($2\times10^{-3}$ relative error), from the original model with 3531 parameters. Moreover, the large model with 106529 parameters finally obtains -0.49707 ($10^{-4}$ relative error) within 3500 iterations.

Similar behaviors are observed for the $16\times16$ system. The final energy comes to -0.4954 ($10^{-3}$ relative error) with 3600 steps for the small model, and the corresponding result is over -0.4961 ($10^{-4}$ relative error) with 1000 steps for the large model. We are confident to draw a conclusion that the increasing parameters significantly strengthen the representation ability for the quantum many-body states from experiments on $10\times10$ and $16\times16$ system.

The comparison results also demonstrate that there is a very limited effect on increasing training steps, where the energy decreased by 0.0001 from the additional 20000 iterations. The major contribution comes from efficient optimization at the early stage. With continuous optimization, the training curve becomes smoother and smoother, and the training becomes more and more difficult.

\subsubsection{Transfer learning on different systems}
Based on the concept of transfer learning and the well-optimized CNN model for the $24\times24$ system,  we append the following fine-tuning optimization on varied systems. For \textbf{Stage 2}, the quantum systems include for the previously discussed scale $10\times10$ and $16\times16$. The model is also used for optimizing the systems for $12\times12$, $14\times14$ and $18\times18$ square lattice, which are never trained before. Moreover, we adjust the value of $J_2/J_1$ from 0.5 to the whole controversial zone from 0.44 to 0.6, in \textbf{Stage 3}.

\begin{table}[bt]
	\centering
	\caption{Energies for varied $J_2/J_1$ \&\& system scales.}
	\resizebox{8.6cm}{!}{
		
		\begin{tabular}{|c|c|c|c|c|c|c|}
			
			\hline
			Description  & 0.44 & 0.46 & 0.48 & 0.50 & 0.55 & 0.60 \\
			\hline
			$10\times10$ & -0.51273 & -0.50728 & -0.50209 & \textbf{-0.49717} & -0.48611 & -0.47713   \\
			$12\times12$ & -0.51225 & -0.50679 & -0.50159 & -0.49665 & -0.48545 & -0.47601   \\
			$14\times14$ & -0.512 & -0.50655 & -0.50135 & -0.4964 & -0.48515 & -0.47558   \\
			$16\times16$ & -0.51188 & -0.50642 & -0.50122 & \textbf{-0.49626} & -0.48499 & -0.47535   \\
			$18\times18$ & -0.51179 & -0.50635 & -0.50114 & -0.49617 & -0.48488 & -0.4752  \\
			\hline
		\end{tabular}
		
	}	
	\label{final-result-table}
	\vspace{-0.3cm}
\end{table}

The final energies obtained by transfer leaning on varied systems are shown in Table \ref {final-result-table}. The correctness of these results has been confirmed from the view of physics, and the relevant physics details will be revealed later. The following content in this section will focus on the effect of transfer learning from the view of computer science.

\begin{figure}[thbp]
	\centerline{\includegraphics[scale=1.1]{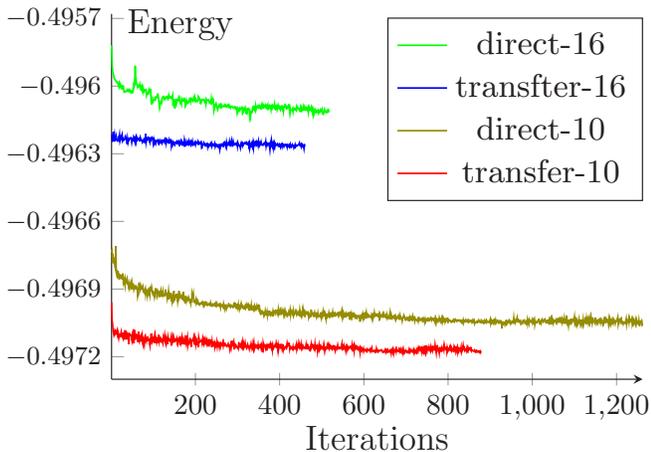}}
	\caption{Comparison on $10\times10$ and $16\times16$ quantum systems with different training process: direct fine-tuning on \textbf{Stage 1} and transfer learning on \textbf{Stage 2}.}
	\label{transferred-training}
	\vspace{-0.1cm}
\end{figure}

As specially marked, there are two results worth noting for $10\times10$ and $16\times16$ system with $J_2/J_1=0.5$. The energies achieved by transfer learning reach -0.49626 and -0.49717, which are far better than the final results from solely \textbf{Stage 1} for -0.49708 and -0.49605. Fig. \ref{transferred-training}  presents the training curves for different kinds of optimization methods: direct training and transfer learning from the $24\times24$ system. The former is purely based on \textbf{Stage 1}, where the CNN model focuses on the target problem and specific system size. The latter is based on \textbf{Stage 2}, which uses the final parameters from the $24\times24$ system.

For the circumstance of the $10\times10$ system, the final energy comes to -0.49705 with 1200 steps for the direct training from system without symmetry, while the optimized model from $24\times24$ system starts at the energy of -0.4969 and easily achieves -0.4971 with initial 20 steps. For the $16\times16$ system, it is almost impossible to obtain an energy value of -0.4961 by increasing the number of iterations. However, the transfer learning solution starts at the energy of -0.4962. In Fig \ref{more-parameters}, the improvement by 0.0001 requires great efforts with considerable computational cost. Consequently, the difference on $10\times10$ system and $16\times16$ system represents significant improvements.

Traditionally, fine-tuning optimization usually requires more accurate data for the target problem, which can be regarded as the most efficient and final stage. However, our experiments demonstrate that the extracted feature from a large-scale system can be significantly applied for improving the result of the original system. Considering that nature is boundless, the actual system is of infinite scale. The larger system is less influenced by the boundary effect, especially for long range correlation effects. By the employment of transfer learning, the isolated problems are connected by practical ties, which brings more characteristics from $24\times24$ system to overcome local minima of $10\times10$ system. Thus, training the model with $24\times24$ system is more efficient than concentrating on smaller system.

\section{Related Work}

In order to optimize the system performance for HPC applications, many researchers apply machine learning methods to explore algorithmic patterns and provide an optimized selection for execution, such as Sparse Matrix Vector Multiplication (SpMV), graph algorithms, and Eulerian fluid simulation\cite{zhao2018bridging,meng2019pattern,li2013smat,dong2019adaptive}. On the other hand, many science areas have made extraordinary breakthroughs by directly developing vigorous deep learning solutions for understanding the relationships between variables in complex data. Especially for exploiting the advantages of modern supercomputers, recent works about climate analytics, physical modeling, and scientific data analysis\cite{jia2020pushing,kurth2018exascale,mathuriya2018cosmoflow}, have revealed great potential for integration of AI and HPC problems on modern supercomputers. Our implementation of the CNN-based wave function representation can be classified into this category. Different from most existing HPC-AI researches, this work demonstrates the effectiveness of mutual consideration on both HPC application requirements and the CNN algorithm characteristics. For one thing, the hybrid output from traditional sign computation and the CNN amplitude is proposed for confronting the sign generation problem. In addition, the variational method is applied over the backpropagation gradients and replaces the prevalent stochastic gradient descent method.

For the quantum many-body problems, the PEPS++ framework\cite{he2018peps++} is firstly proposed as a computational method towards extreme-scale simulations on Sunway TaihuLight. Recent study\cite{pang2020efficient} with distributed 2D tensor further extends efficiently two-layer PEPS contraction by distributed-memory tensor computation for approximate simulations on $J_1-J_2$ model with $4\times4$ quantum system on Stampede2 supercomputer, which differs from our proposal. Another relevant research\cite{nomura2021dirac} is delivered from RIKEN's group, which performs a systematic investigation for no more than $18\times18$ quantum system on the K/Fugaku supercomputer. It achieves state-of-the-art accuracy and their ground state energies are chosen for benchmarks. Meanwhile, another preliminary preparation is about constructing the deep learning ecosystem on sunway supercomputer\cite{li2021swflow}, which presents systematic optimizations on kernel acceleration, communication optimization.

The last point worth mentioning is about model optimization. The remarkable works\cite{you2017large, you2018imagenet,you2019large}, like Layer-wise Adaptive Rate Scaling and linear-epoch gradual-warmup, are widely applied for scaling model training in data parallelism. However, it is unfit for quantum systems, which involve second derivatives statistics data by SR method\cite{casula2005diffusion,carleo2017solving}. More relevant works\cite{gupta2018shampoo, martens2015optimizing, anil2020scalable, osawa2019large} adopt second-order optimization by building the covariance matrix to accumulate the outer products of stochastic gradients. For confronting the prohibitive runtime and memory costs, they adopt the Kronecker product to approximate the preconditioning matrix (as dimension can be as large as $10^4$). The handling of the covariance matrix in this work is implemented by the distributed ScaLapack library, while it limits the scalability for matrix scale and the number of variational parameters.

From the perspective of sunway supercomputer, there are numerous studies on achieving high performance scientific computing with a series of optimization techniques. Recently, a tensor network based quantum simulator (SW\_Qsim) was introduced, which simulated the random quantum circuit with up to 400-qubits.\cite{li2021sw_qsim}. Compared to our optimization problem, this application only requires random sampling in quantum space, which requires little communication and thus achieves near-linear scaling with up to 28.75 million cores. A deep learning driven kinetic Monte Carlo simulation application\cite{shang2021tensorkmc}, TensorKMC, integrates neural network with classic physical simulations, which adopts the network inference for generating sampling data. It differs from our application
that it only uses the neural network inference/forward for generating data samples, whereas our application further needs model training/backward. This application also obtains excellent scalability on 422400 CGs with over 27 million cores. In addition, other traditional high performance applications are also implemented on sunway supercomputer, including solving sparse linear system equations, molecular dynamics, and computational fluid dynamics\cite{zhu2021enabling,gao2021lmff,huang2019heterogeneous,li2020memory}. Especially, the HPCG can be regarded as a benchmark for supercomputer system, which finally achieves 95.5\% scaling efficiency on 42 million heterogeneous cores\cite{zhu2021enabling}. One of the LAMMPS based implementations, layered materials force field (LMFF), is evaluated with parallel efficiency of 88\% on 196608 processes\cite{zhu2021enabling}, which achieves great progress over the original applications on the Sunway TaihuLight\cite{huang2019heterogeneous,li2020memory,lin2019swflow}. Compared to the aforementioned  studies\cite{li2021sw_qsim,shang2021tensorkmc}, the current work is unique that it involves both Monte Carlo data sampling and model training to obtain the ground state of a quantum many-body systems. The distributed SR optimization consequently increases the communication overhead. However, this application still reaches competitive parallel efficiency of 92.5\% for 490000 processes of 31.85 million cores.

Overall, this paper is developed from two previous papers for solving the spin-$1/2$ $J_1-J_2$ model with CNN-based ansatz combining the latest progress from HPC and AI. Based on the magnificent computing power of the supercomputer, the model parameter is increased by 30x from 3531 to 106529. Compared to relevant influential works\cite{sorella2007weak, choo2019two, westerhout2020generalization} with thousands of parameters, it achieves great improvement by an order of magnitude in system scale. Most importantly, the multiple stages of transfer learning greatly reduce the computational cost and firstly extend the quantum system scale as $24\times24$ system. Compared to the state-of-the-art results on the lattice size of $18\times18$, the search space of this work is $2^{252}$ times larger, which has revealed great potential for integration of AI and HPC advantages.

\section{Conclusion Remarks}
The efficient simulation of quantum many-body problem plays a key role in understanding the exotic physical properties in condensed matter physics. In this work, we present a novel low scaling cost and high scalable CNN-based wave function representation methods for handling the exponential complexity of representing quantum many-body states. We demonstrated our method on the 2D strongly frustrated spin-$1/2$ $J_1-J_2$ model and showed that it can manage millions of independent Markov chains with over 30 million cores of the new sunway supercomputer. By ingenious transfer learning, we investigate systems with size up to $24\times24$ and bring the state-of-art record up to a brand new level from both aspects of remarkable accuracy and unprecedented scales.

To take full potential of heterogeneous sw26010pro processor, we exploited the intrinsic parallelism for inter- and intra-MPI processes by optimizing the sequential serial sampling process. The most time-consuming computation of the CNN is then offloaded onto the CPE cluster through swDNN library. As for the strict limitation to the number of variational parameters, the tricky computation for the covariance matrix from SR method is addressed by the parallel and distributed ScaLapack library, as well as better alleviation of the computational cost. Finally, transfer learning not only simplifies the task difficulty and reduces the computational cost for large-scale quantum systems, but opens up a promising route for confronting the challenge at unprecedented scales.

With the great success on the quantum many-body problems for spin-$1/2$ $J_1-J_2$ model in this study, the CNN-based solution can be further applied to other challenging problems. Meanwhile, this work demonstrates the effectiveness of collaborative considerations for HPC application requirements and CNN algorithm characteristics, as well as providing a promising paradigm and good reference for future scenarios.

\ifCLASSOPTIONcompsoc
\section*{Acknowledgments}
\else
\section*{Acknowledgment}
\fi

We sincerely thank the editors and the reviewers for their careful reading and thoughtful comments. The work is supported by the National Key Research and Development Program of China (Grant No. 2016YFB1000403).

\ifCLASSOPTIONcaptionsoff
\newpage
\fi


\bibliographystyle{IEEEtran}
\bibliography{IEEEabrv,swtf}

\begin{thebibliography}{10}
\providecommand{\url}[1]{#1}
\csname url@samestyle\endcsname
\providecommand{\newblock}{\relax}
\providecommand{\bibinfo}[2]{#2}
\providecommand{\BIBentrySTDinterwordspacing}{\spaceskip=0pt\relax}
\providecommand{\BIBentryALTinterwordstretchfactor}{4}
\providecommand{\BIBentryALTinterwordspacing}{\spaceskip=\fontdimen2\font plus
\BIBentryALTinterwordstretchfactor\fontdimen3\font minus
  \fontdimen4\font\relax}
\providecommand{\BIBforeignlanguage}[2]{{%
\expandafter\ifx\csname l@#1\endcsname\relax
\typeout{** WARNING: IEEEtran.bst: No hyphenation pattern has been}%
\typeout{** loaded for the language `#1'. Using the pattern for}%
\typeout{** the default language instead.}%
\else
\language=\csname l@#1\endcsname
\fi
#2}}
\providecommand{\BIBdecl}{\relax}
\BIBdecl

\bibitem{kohn1999nobel}
W.~Kohn, ``Nobel lecture: Electronic structure of matter—wave functions and
  density functionals,'' \emph{Reviews of Modern Physics}, vol.~71, no.~5, p.
  1253, 1999.

\bibitem{gygi2006large}
F.~Gygi, E.~W. Draeger, M.~Schulz, B.~R. De~Supinski, J.~A. Gunnels, V.~Austel,
  J.~C. Sexton, F.~Franchetti, S.~Kral, C.~W. Ueberhuber \emph{et~al.},
  ``Large-scale electronic structure calculations of high-z metals on the
  bluegene/l platform,'' in \emph{Proceedings of the 2006 ACM/IEEE conference
  on Supercomputing}, 2006, pp. 45--es.

\bibitem{hasegawa2011first}
Y.~Hasegawa, J.-I. Iwata, M.~Tsuji, D.~Takahashi, A.~Oshiyama, K.~Minami,
  T.~Boku, F.~Shoji, A.~Uno, M.~Kurokawa \emph{et~al.}, ``First-principles
  calculations of electron states of a silicon nanowire with 100,000 atoms on
  the k computer,'' in \emph{Proceedings of 2011 International Conference for
  High Performance Computing, Networking, Storage and Analysis}, 2011, pp.
  1--11.

\bibitem{carleo2017solving}
G.~Carleo and M.~Troyer, ``Solving the quantum many-body problem with
  artificial neural networks,'' \emph{Science}, vol. 355, no. 6325, pp.
  602--606, 2017.

\bibitem{you2017large}
Y.~You, I.~Gitman, and B.~Ginsburg, ``Large batch training of convolutional
  networks,'' \emph{arXiv preprint arXiv:1708.03888}, 2017.

\bibitem{you2018imagenet}
Y.~You, Z.~Zhang, C.-J. Hsieh, J.~Demmel, and K.~Keutzer, ``Imagenet training
  in minutes,'' in \emph{Proceedings of the 47th International Conference on
  Parallel Processing}, 2018, pp. 1--10.

\bibitem{you2019large}
Y.~You, J.~Hseu, C.~Ying, J.~Demmel, K.~Keutzer, and C.-J. Hsieh, ``Large-batch
  training for lstm and beyond,'' in \emph{Proceedings of the International
  Conference for High Performance Computing, Networking, Storage and Analysis},
  2019, pp. 1--16.

\bibitem{neuscamman2012optimizing}
E.~Neuscamman, C.~Umrigar, and G.~K.-L. Chan, ``Optimizing large parameter sets
  in variational quantum monte carlo,'' \emph{Physical Review B}, vol.~85,
  no.~4, p. 045103, 2012.

\bibitem{sorella1998green}
S.~Sorella, ``Green function monte carlo with stochastic reconfiguration,''
  \emph{Physical review letters}, vol.~80, no.~20, p. 4558, 1998.

\bibitem{benedict2019quantum}
K.~A. Benedict, ``Quantum monte carlo methods: algorithms for lattice models,''
  \emph{Contemporary Physics}, vol.~60, no.~1, pp. 80--81, 2019.

\bibitem{liang2018solving}
X.~Liang, W.-Y. Liu, P.-Z. Lin, G.-C. Guo, Y.-S. Zhang, and L.~He, ``Solving
  frustrated quantum many-particle models with convolutional neural networks,''
  \emph{Physical Review B}, vol.~98, no.~10, p. 104426, 2018.

\bibitem{liang2021hybrid}
X.~Liang, S.-J. Dong, and L.~He, ``Hybrid convolutional neural network and
  projected entangled pair states wave functions for quantum many-particle
  states,'' \emph{Physical Review B}, vol. 103, no.~3, p. 035138, 2021.

\bibitem{nomura2021dirac}
Y.~Nomura and M.~Imada, ``Dirac-type nodal spin liquid revealed by refined
  quantum many-body solver using neural-network wave function, correlation
  ratio, and level spectroscopy,'' \emph{Physical Review X}, vol.~11, no.~3, p.
  031034, 2021.

\bibitem{choi1992scalapack}
J.~Choi, J.~J. Dongarra, R.~Pozo, and D.~W. Walker, ``Scalapack: A scalable
  linear algebra library for distributed memory concurrent computers,'' in
  \emph{The Fourth Symposium on the Frontiers of Massively Parallel
  Computation}.\hskip 1em plus 0.5em minus 0.4em\relax IEEE Computer Society,
  1992, pp. 120--121.

\bibitem{voigt2000marshall}
A.~Voigt and J.~Richter, ``Marshall-peierls sign rule in frustrated heisenberg
  chains,'' \emph{arXiv preprint cond-mat/0003207}, 2000.

\bibitem{szabo2020neural}
A.~Szab{\'o} and C.~Castelnovo, ``Neural network wave functions and the sign
  problem,'' \emph{Physical Review Research}, vol.~2, no.~3, p. 033075, 2020.

\bibitem{fu2016sunway}
H.~Fu, J.~Liao, J.~Yang, L.~Wang, Z.~Song, X.~Huang, C.~Yang, W.~Xue, F.~Liu,
  F.~Qiao \emph{et~al.}, ``The sunway taihulight supercomputer: system and
  applications,'' \emph{Science China Information Sciences}, vol.~59, no.~7,
  pp. 1--16, 2016.

\bibitem{dongarra2016sunway}
J.~Dongarra, ``Sunway taihulight supercomputer makes its appearance,''
  \emph{National Science Review}, vol.~3, no.~3, pp. 265--266, 2016.

\bibitem{gao2021optimization}
P.~Gao, X.~Duan, B.~Schmidt, W.~Zhang, L.~Gan, H.~Fu, W.~Xue, W.~Liu, and
  G.~Yang, ``Optimization of reactive force field simulation: Refactor,
  parallelization, and vectorization for interactions,'' \emph{IEEE
  Transactions on Parallel and Distributed Systems}, 2021.

\bibitem{geyer1992practical}
C.~J. Geyer, ``Practical markov chain monte carlo,'' \emph{Statistical
  science}, pp. 473--483, 1992.

\bibitem{li2021swflow}
M.~Li, H.~Lin, J.~Chen, J.~M. Diaz, Q.~Xiao, R.~Lin, F.~Wang, G.~R. Gao, and
  H.~An, ``swflow: A large-scale distributed framework for deep learning on
  sunway taihulight supercomputer,'' \emph{Information Sciences}, vol. 570, pp.
  831--847, 2021.

\bibitem{kim2008chiral}
J.~H. Kim and J.~H. Han, ``Chiral spin states in the pyrochlore heisenberg
  magnet: Fermionic mean-field theory and variational monte carlo
  calculations,'' \emph{Physical Review B}, vol.~78, no.~18, p. 180410, 2008.

\bibitem{fang2017swdnn}
J.~Fang, H.~Fu, W.~Zhao, B.~Chen, W.~Zheng, and G.~Yang, ``swdnn: A library for
  accelerating deep learning applications on sunway taihulight,'' in \emph{2017
  IEEE International Parallel and Distributed Processing Symposium
  (IPDPS)}.\hskip 1em plus 0.5em minus 0.4em\relax IEEE, 2017, pp. 615--624.

\bibitem{unpublished}
X.~Liang \emph{et~al.}, ``unpublished paper,'' 2021.

\bibitem{zhao2018bridging}
Y.~Zhao, J.~Li, C.~Liao, and X.~Shen, ``Bridging the gap between deep learning
  and sparse matrix format selection,'' in \emph{Proceedings of the 23rd ACM
  SIGPLAN symposium on principles and practice of parallel programming}, 2018,
  pp. 94--108.

\bibitem{meng2019pattern}
K.~Meng, J.~Li, G.~Tan, and N.~Sun, ``A pattern based algorithmic autotuner for
  graph processing on gpus,'' in \emph{Proceedings of the 24th Symposium on
  Principles and Practice of Parallel Programming}, 2019, pp. 201--213.

\bibitem{li2013smat}
J.~Li, G.~Tan, M.~Chen, and N.~Sun, ``Smat: an input adaptive auto-tuner for
  sparse matrix-vector multiplication,'' in \emph{Proceedings of the 34th ACM
  SIGPLAN conference on Programming language design and implementation}, 2013,
  pp. 117--126.

\bibitem{dong2019adaptive}
W.~Dong, J.~Liu, Z.~Xie, and D.~Li, ``Adaptive neural network-based
  approximation to accelerate eulerian fluid simulation,'' in \emph{Proceedings
  of the International Conference for High Performance Computing, Networking,
  Storage and Analysis}, 2019, pp. 1--22.

\bibitem{jia2020pushing}
W.~Jia, H.~Wang, M.~Chen, D.~Lu, L.~Lin, R.~Car, E.~Weinan, and L.~Zhang,
  ``Pushing the limit of molecular dynamics with ab initio accuracy to 100
  million atoms with machine learning,'' in \emph{SC20: International
  Conference for High Performance Computing, Networking, Storage and
  Analysis}.\hskip 1em plus 0.5em minus 0.4em\relax IEEE, 2020, pp. 1--14.

\bibitem{kurth2018exascale}
T.~Kurth, S.~Treichler, J.~Romero, M.~Mudigonda, N.~Luehr, E.~Phillips,
  A.~Mahesh, M.~Matheson, J.~Deslippe, M.~Fatica \emph{et~al.}, ``Exascale deep
  learning for climate analytics,'' in \emph{SC18: International Conference for
  High Performance Computing, Networking, Storage and Analysis}.\hskip 1em plus
  0.5em minus 0.4em\relax IEEE, 2018, pp. 649--660.

\bibitem{mathuriya2018cosmoflow}
A.~Mathuriya, D.~Bard, P.~Mendygral, L.~Meadows, J.~Arnemann, L.~Shao, S.~He,
  T.~K{\"a}rn{\"a}, D.~Moise, S.~J. Pennycook \emph{et~al.}, ``Cosmoflow: Using
  deep learning to learn the universe at scale,'' in \emph{SC18: International
  Conference for High Performance Computing, Networking, Storage and
  Analysis}.\hskip 1em plus 0.5em minus 0.4em\relax IEEE, 2018, pp. 819--829.

\bibitem{he2018peps++}
L.~He, H.~An, C.~Yang, F.~Wang, J.~Chen, C.~Wang, W.~Liang, S.~Dong, Q.~Sun,
  W.~Han \emph{et~al.}, ``Peps++: towards extreme-scale simulations of strongly
  correlated quantum many-particle models on sunway taihulight,'' \emph{IEEE
  Transactions on Parallel and Distributed Systems}, vol.~29, no.~12, pp.
  2838--2848, 2018.

\bibitem{pang2020efficient}
Y.~Pang, T.~Hao, A.~Dugad, Y.~Zhou, and E.~Solomonik, ``Efficient 2d tensor
  network simulation of quantum systems,'' in \emph{SC20: International
  Conference for High Performance Computing, Networking, Storage and
  Analysis}.\hskip 1em plus 0.5em minus 0.4em\relax IEEE, 2020, pp. 1--14.

\bibitem{casula2005diffusion}
M.~Casula, C.~Filippi, and S.~Sorella, ``Diffusion monte carlo method with
  lattice regularization,'' \emph{Physical review letters}, vol.~95, no.~10, p.
  100201, 2005.

\bibitem{gupta2018shampoo}
V.~Gupta, T.~Koren, and Y.~Singer, ``Shampoo: Preconditioned stochastic tensor
  optimization,'' in \emph{International Conference on Machine Learning}.\hskip
  1em plus 0.5em minus 0.4em\relax PMLR, 2018, pp. 1842--1850.

\bibitem{martens2015optimizing}
J.~Martens and R.~Grosse, ``Optimizing neural networks with kronecker-factored
  approximate curvature,'' in \emph{International conference on machine
  learning}.\hskip 1em plus 0.5em minus 0.4em\relax PMLR, 2015, pp. 2408--2417.

\bibitem{anil2020scalable}
R.~Anil, V.~Gupta, T.~Koren, K.~Regan, and Y.~Singer, ``Scalable second order
  optimization for deep learning,'' \emph{arXiv preprint arXiv:2002.09018},
  2020.

\bibitem{osawa2019large}
K.~Osawa, Y.~Tsuji, Y.~Ueno, A.~Naruse, R.~Yokota, and S.~Matsuoka,
  ``Large-scale distributed second-order optimization using kronecker-factored
  approximate curvature for deep convolutional neural networks,'' in
  \emph{Proceedings of the IEEE/CVF Conference on Computer Vision and Pattern
  Recognition}, 2019, pp. 12\,359--12\,367.

\bibitem{li2021sw_qsim}
F.~Li, X.~Liu, Y.~Liu, P.~Zhao, Y.~Yang, H.~Shang, W.~Sun, Z.~Wang, E.~Dong,
  and D.~Chen, ``Sw\_qsim: a minimize-memory quantum simulator with
  high-performance on a new sunway supercomputer,'' in \emph{Proceedings of the
  International Conference for High Performance Computing, Networking, Storage
  and Analysis}, 2021, pp. 1--13.

\bibitem{shang2021tensorkmc}
H.~Shang, X.~Chen, X.~Gao, R.~Lin, L.~Wang, F.~Li, Q.~Xiao, L.~Xu, Q.~Sun,
  L.~Zhu \emph{et~al.}, ``Tensorkmc: kinetic monte carlo simulation of 50
  trillion atoms driven by deep learning on a new generation of sunway
  supercomputer,'' in \emph{Proceedings of the International Conference for
  High Performance Computing, Networking, Storage and Analysis}, 2021, pp.
  1--14.

\bibitem{zhu2021enabling}
Q.~Zhu, H.~Luo, C.~Yang, M.~Ding, W.~Yin, and X.~Yuan, ``Enabling and scaling
  the hpcg benchmark on the newest generation sunway supercomputer with 42
  million heterogeneous cores,'' in \emph{Proceedings of the International
  Conference for High Performance Computing, Networking, Storage and Analysis},
  2021, pp. 1--13.

\bibitem{gao2021lmff}
P.~Gao, X.~Duan, J.~Guo, J.~Wang, Z.~Song, L.~Cui, X.~Meng, X.~Liu, W.~Zhang,
  M.~Ma \emph{et~al.}, ``Lmff: efficient and scalable layered materials force
  field on heterogeneous many-core processors,'' in \emph{Proceedings of the
  International Conference for High Performance Computing, Networking, Storage
  and Analysis}, 2021, pp. 1--14.

\bibitem{huang2019heterogeneous}
J.~Huang, W.~Han, X.~Wang, and W.~Chen, ``Heterogeneous parallel algorithm
  design and performance optimization for weno on the sunway taihulight
  supercomputer,'' \emph{Tsinghua Science and Technology}, vol.~25, no.~1, pp.
  56--67, 2019.

\bibitem{li2020memory}
J.~Li, J.~Lin, P.~Du, K.~Zhang, and J.~Wu, ``Memory access optimization of
  molecular dynamics simulation software crystal-md on sunway taihulight,''
  \emph{Tsinghua Science and Technology}, vol.~26, no.~3, pp. 296--308, 2020.

\bibitem{lin2019swflow}
H.~Lin, Z.~Lin, J.~M. Diaz, M.~Li, H.~An, and G.~R. Gao, ``swflow: a dataflow
  deep learning framework on sunway taihulight supercomputer,'' in \emph{2019
  IEEE 21st International Conference on High Performance Computing and
  Communications; IEEE 17th International Conference on Smart City; IEEE 5th
  International Conference on Data Science and Systems
  (HPCC/SmartCity/DSS)}.\hskip 1em plus 0.5em minus 0.4em\relax IEEE, 2019, pp.
  2467--2475.

\bibitem{sorella2007weak}
S.~Sorella, M.~Casula, and D.~Rocca, ``Weak binding between two aromatic rings:
  Feeling the van der waals attraction by quantum monte carlo methods,''
  \emph{The Journal of chemical physics}, vol. 127, no.~1, p. 014105, 2007.

\bibitem{choo2019two}
K.~Choo, T.~Neupert, and G.~Carleo, ``Two-dimensional frustrated j 1- j 2 model
  studied with neural network quantum states,'' \emph{Physical Review B}, vol.
  100, no.~12, p. 125124, 2019.

\bibitem{westerhout2020generalization}
T.~Westerhout, N.~Astrakhantsev, K.~S. Tikhonov, M.~I. Katsnelson, and A.~A.
  Bagrov, ``Generalization properties of neural network approximations to
  frustrated magnet ground states,'' \emph{Nature communications}, vol.~11,
  no.~1, pp. 1--8, 2020.

\end{thebibliography}

\vspace{12pt}



%

\end{document}